\documentclass[3p,times,twocolumn]{elsarticle}

\usepackage{amssymb}

\usepackage[figuresright]{rotating}

\newcommand{\bc}{\begin{center}}
\newcommand{\ec}{\end{center}}
\newcommand{\be}{\begin{equation}}
\newcommand{\ee}{\end{equation}}
\newcommand{\bea}{\begin{eqnarray}}
\newcommand{\eea}{\end{eqnarray}}
\newcommand{\ba}{\begin{array}}
\newcommand{\ea}{\end{array}}
\newcommand{\lb}{\label}
\newcommand{\rf}{\ref}
\newcommand{\bfg}{\begin{figure}[htbp]}
\newcommand{\efg}{\end{figure}}

\begin{document}

\begin{frontmatter}




\title{On the role of dynamical quark mass generation \protect \\
in chiral symmetry breaking in QCD}


\author{H. Sazdjian}

\address{Institut de Physique Nucl\'eaire, CNRS/IN2P3,\\
Universit\'e Paris-Sud, F-91405 Orsay, France}

\ead{sazdjian@ipno.in2p3.fr}

\begin{abstract}
The phenomenon of dynamical quark mass generation is studied in 
QCD within the framework of a gauge invariant formalism. An exact 
relationship is established between the equation satisfied by the 
scalar part of the two-point gauge invariant quark Green's function
and the quark-antiquark bound state equation in the chiral limit. 
A possible nontrivial solution of the former yields a massless 
pseudoscalar solution of the bound state equation with vanishing total 
momentum. The result is also corroborated by the corresponding 
Ward-Takahashi identity. The problem is explicitly solved in 
two-dimensional QCD in the large-$N_c$ limit.       
\end{abstract}


\end{frontmatter}


\section{Introduction} \lb{s1}

It has been known for a long time, since the advent of the 
Nambu--Jona-Lasinio model \cite{Nambu:1961tp}, that 
a mechanism of chiral symmetry breaking in theories with fermions
as the elementary matter fields is provided by the dynamical mass 
generation phenomenon. This mechanism was also analyzed in the 
framework of QED by Baker, Johnson and Lee \cite{Baker:1964zza},
but the presence of the axial-vector current anomaly in Abelian
sectors of the chiral group did not allow drawing a similar 
conclusion. 
\par
The problem was later studied in the framework of QCD, with
the aid of models applied to the Dyson--Schwinger equations 
\cite{Dyson:1949ha,Schwinger:1951ex,Alkofer:2000wg,
Fischer:2006ub} in the Coulomb gauge, where the gluon propagator in 
the rainbow-ladder approximation is considered with its instantaneous 
expression, with an appropriate confining behavior 
\cite{Finger:1980dw,LeYaouanc:1983iy,Adler:1984ri,Alkofer:1988tc}. 
The dynamical mass generation phenomenon was confirmed
and phenomenological applications to the meson spectrum were considered.
An extension of the approximation beyond the rainbow-ladder one 
preserved the stability of the phenomenon \cite{Bender:1996bb}.
On the other hand, the use of the Ward-Takahashi identity for the 
axial-vector current \cite{Preparata:1969hg} establishes a general 
relationship, independent of approximation schemes of the Bethe--Salpeter 
kernel, between the dynamically generated quark mass and the pseudoscalar
Goldstone boson wave function \cite{Maris:1997hd}.   
\par
In spite of these positive results, the notion of dynamical quark mass 
remains ambiguous. Contrary to the case of nonconfining theories, the
quark mass is not an observable quantity. Furthermore, the quark 
propagator not being a gauge invariant object, it is difficult to 
give a physical interpretation to its singularities. In many examples,
an infrared cutoff is needed to make it finite.
\par
The aim of this talk is to present results obtained within a gauge
invariant formalism, developed in recent years \cite{Sazdjian:2007ng,
Sazdjian:2013rva}, where an exact relationship is established between
the dynamical quark mass term and the pseudoscalar Goldstone boson
wave function in the total zero momentum limit. 
The explicit resolution of the problem in analytic form in 
two-dimensional QCD in the large-$N_c$ limit illustrates the advantage
of the gauge invariant approach in providing a physical insight into
the interpretation of the dynamical mass generation phenomenon.
\par

\section{Gauge invariant quark Green's functions} \lb{s2}

To define gauge invariant quark Green's functions, one needs to
introduce gluon field path-ordered phase factors (Wilson lines)
\cite{Mandelstam:1968hz,Nambu:1978bd}, which are the operators of 
parallel transport of fields from one point to the other. The paths 
followed by the phase factors might be arbitrarily chosen, however, for 
the quark Green's functions polygonal lines seem to correspond to the 
optimal choice. First, they can be decomposed along a succession of 
straight line segments. The latter are Lorentz invariant in form and 
this in turn allows the classification of the polygonal lines 
according to the number of segments they contain. Second, polygonal 
lines form a complete set of lines for the description of the present 
problem, in the sense that no other types of line are needed to complete 
the description.
\par
The classification adopted for the polygonal lines is applied 
to the two-point gauge invariant quark Green's functions (2PGIQGF).
We designate by $U(y,x)$ the path-ordered phase factor along
the oriented straight line segment going from $x$ to $y$.
The quark fields, with mass parameter $m$, are assumed to belong 
to the defining fundamental ($N_c$-dimensional) representation of 
the color gauge group $SU(N_c)$. The 2PGIQGF with a polygonal line 
composed of the succession of $n$ segments with $(n-1)$ junction 
points is designated by $S_{(n)}$ and defined as
\bea \lb{2e1}
\lefteqn{S_{(n)}(x,x';t_{n-1},\ldots,t_1)=-\frac{1}{N_c}\langle 
\overline\psi(x')U(x',t_{n-1})}\nonumber \\
& &\ \ \ \ \ \ \ \times U(t_{n-1},t_{n-2})\ldots U(t_1,x)\psi(x)\rangle.
\eea
(The vacuum expectation value is defined in the path integral 
formalism; spinor indices are omitted and the color indices are 
implicitly summed.) The simplest such function is $S_{(1)}$, having 
a phase factor along one straight line segment:
\be \lb{2e2} 
S_{(1)}(x,x')=-\frac{1}{N_c}\langle 
\overline\psi(x')U(x',x)\psi(x)\rangle.
\ee
\par
Use of the equations of motion of Green's functions and integrations
yield functional relations between the Green's functions of different
classes of polygonal line. It turns out that $S_{(1)}$ is the only
dynamically independent 2PGIQGF. All other 2PGIQGFs $S_{(n)}$, with
$n>1$, are calculable from $S_{(1)}$ through the functional relations.
These involve essentially Wilson loops \cite{Wilson:1974sk} along
polygonal contours and functional derivatives acting on their sides.
\par   
The equation satisfied by $S_{(1)}$ can be brought to the following
form:
\bea \lb{2e3}
\lefteqn{\hspace{-0.5 cm}(i\gamma.\partial_{(x)}-m)\,S_{(1)}(x,x')
=i\delta^4(x-x')+i\gamma^{\mu}\Big\{K_{1\mu}(x',x)}\nonumber \\
& &\times S_{(1)}(x,x')
+K_{2\mu}(x',x,y_1)\,S_{(2)}(y_1,x';x)\nonumber \\
& &+\sum_{n=3}^{\infty}K_{n\mu}(x',x,y_1,\ldots,y_{n-1})
\nonumber \\
& &\ \ \ \ \ \ \ \ \times S_{(n)}(y_{n-1},x';x,y_1,\ldots,y_{n-2})\Big\},
\eea
where the kernels $K_n$ ($n=1,2,\ldots$) contain Wilson loop averages
along polygonal contours, which are at most $(n+1)$-sided, and ($n-1$) 
2PGIQGFs $S_{(1)}$ and their derivative. The total number of derivatives 
contained in $K_n$ is $n$, each derivative acting on a different segment.
Since the high-index $S_{(n)}$s can be expressed in terms
of $S_{(1)}$, Eq. (\rf{2e3}) is an integro\-differen\-tial equation 
in $S_{(1)}$, which is the primary unknown quantity to be solved.
(Integration symbols on the intermediate variables have been omitted
in the right-hand side.) Equation (\rf{2e3}) is the analog of the
self-energy Dyson--Schwinger equation in the present formalism. 
\par

\section{Bound state wave functions} \lb{s3}

The previous definitions and approach can also be applied to 
four-point Green's functions. We introduce two quark fields (in
flavor space) $\psi_1$ and $\psi_2$, with masses $m_1$ and $m_2$,
respectively, and consider gauge invariant four-point Green's
functions made of two quark and two antiquark fields and 
appropriate phase factors along polygonal lines. Considering a
mesonic bound state, with total momentum $P$, one defines 
corresponding gauge invariant wave functions with a similar
classification as for the 2PGIQGFs ($n=1,2,\ldots$):
\bea \lb{3e1}
\lefteqn{\hspace{-0.5 cm}\Phi_{(n)}(P;x_1,x_2;t_{n-1},t_{n-2},
\ldots,t_1)=}\nonumber \\
& &-\frac{1}{\sqrt{N_c}}<0|\ T(\overline \psi_{2}(x_2)
U(x_2,t_{n-1})\nonumber \\
& &\times U(t_{n-1},t_{n-2})\ldots U(t_1,x_1)\psi_{1}(x_1))\ |P>.
\eea
In the simplest case $n=1$, one has
\bea \lb{3e2}
\lefteqn{\hspace{-0.5 cm}\Phi_{(1)}(P;x_1,x_2)=}\nonumber \\
\lefteqn{\hspace{-0.2 cm}-\frac{1}{\sqrt{N_c}}
<0|\ T(\overline \psi_2(x_2)U(x_2,x_1)\psi_1(x_1))\ |P>.}
\eea
The above wave functions, for all $n$s, describe the same bound state, 
but differ in their expressions due to their differences in their 
contents with respect to the phase factor lines.   
\par
As in the case of the 2PGIQGFs, functional relations exist between
the above wave functions: $\Phi_{(n)}$s, with $n>1$, are calculable
in terms of $\Phi_{(1)}$.
\par
The wave function $\Phi_{(1)}$ satisfies two Dirac-type 
integro\-diffe\-ren\-tial equations, which can be written in the 
following compact forms:
\bea 
\lb{3e3}
\lefteqn{\hspace{-0.5 cm}(i\gamma.\partial_1-m_1)\Phi_{(1)}(P;x_1,x_2)
=+i\gamma^{\mu}\,\Big\{\sum_{n=1}^{\infty}K_{1,n\mu}*\Phi_{(n)}}
\nonumber \\
& &\ \ \ +\ \sum_{n=1}^{\infty}(N_{1,n\mu}*\Phi_{(1)})*S_{2(n)}\Big\},
\\
\lb{3e4}
\lefteqn{\hspace{-0.5 cm}\Phi_{(1)}(P;x_1,x_2)
(-i\gamma.\stackrel{\leftarrow}{\partial}_2-m_2)
=-i\Big\{\sum_{n=1}^{\infty}\Phi_{(n)}*K_{2,n\nu}}\nonumber \\
& &\ \ \ +\ \sum_{n=1}^{\infty}S_{1(n)}*(\Phi_{(1)}*N_{2,n\nu})\Big\}
\,\gamma^{\nu}.
\eea
The kernels $K_n$ are similar to those met in Eq. (\rf{2e3}); they 
contain Wilson loop averages along polygonal contours with $(n+1)$
sides; the additional indices 1 or 2 reflect the differences that
arise from the presence there of the $(n-1)$ 2PGIQGFs $S_{(1)}$ of 
quarks 1 and/or 2. The kernels $N_n$, with indices 1 or 2, are 
obtained from the $K_n$s by functional derivation with respect to 
one of the 2PGIQGFs. The star operations are compact notaions for the 
various integration operations on the intermediate variables. Since 
the wave functions $\Phi_{(n)}$ are calculable in terms of $\Phi_{(1)}$, 
the above equations are ultimately wave equations for $\Phi_{(1)}$.
\par 
 
\section{Chiral symmetry breaking} \lb{s4}

To analyze the chiral symmetry breaking possibility, we consider 
the 2PGIQGF $S_{(1)}(x_1,x_2)$. Since the single straight line segment 
that is present in the path-ordered phase factor depends only on the end 
points $x_1$ and $x_2$, we can decompose $S_{(1)}$ along two Poincar\'e
invariant parts, made of a vector and of a scalar:
\be \lb{4e1}
S_{(1)}(x)=i\gamma.\partial F_1(x^2)+F_0(x^2).
\ee
($x=x_1-x_2$.) In perturbation theory, the scalar component $F_0$ is
proportional to the quark mass $m$. When the latter vanishes, $F_0$
also vanishes. Therefore, if the solution of the equation of $S_{(1)}$
yields a scalar part that does not vanish in the limit of a massless
quark, then one is in the presence of a non\-pertur\-bative solution,
which may be described as corresponding to the phenomenon of dynamical 
mass generation. This mass term might have a complicated structure,
neither being a constant nor producing a simple pole in the 2PGIQGF.   
\par
The equation satisfied by $F_0$ can be extracted from that of $S_{(1)}$
[Eq. (\rf{2e3})] by considering the anticommutator of $\gamma_5$ with
$S_{(1)}$. One obtains in the massless quark limit
\bea \lb{4e2}
\lefteqn{\hspace{-0.5 cm}i\gamma.\partial_1
[\gamma_5,S_{(1)}(x_1,x_2)]_+^{}=
i\gamma^{\mu}\Big\{\sum_{n=1}^{\infty}K_{n\mu}*[\gamma_5,S_{(n)}]_+^{}}
\nonumber \\
& &\ \ \ \ +\ \sum_{n=1}^{\infty}(N_{n\mu}*[\gamma_5,S_{(1)}]_+^{})
*S_{(n)}\Big\}.
\eea
\par
This equation is to be compared with that of the wave function 
$\Phi_{(1)}$ in the massless limit:
\bea \lb{4e3}
\lefteqn{\hspace{-0.5 cm}i\gamma.\partial_1\Phi_{(1)}(P;x_1,x_2)=
i\gamma^{\mu}\Big\{\sum_{n=1}^{\infty}K_{n\mu}*\Phi_{(n)}}
\nonumber \\
& &\ \ \ \ +\ \sum_{n=1}^{\infty}(N_{n\mu}*\Phi_{(1)})*S_{(n)}\Big\}.
\eea
(The quark indices 1 and 2 have been removed, since the quarks are 
massless.) Equations similar to (\rf{4e2}) and (\rf{4e3}) with
operators acting from the right on the variable $x_2$ could also be 
written down.
\par
We observe that the equation of $[\gamma_5,S_{(1)}]_+^{}$ is the 
same as the equation of $\Phi_{1)}$, provided the general
correspondences $[\gamma_5,S_{(n)}]_+^{}\longrightarrow \Phi_{(n)}$
are also made. Actually, these correspondences are independently
justified by comparing the equations each of them satisfy. They represent
the general consistency conditions for the similarity of the set of
all equations of the above types.
\par
Therefore, if the equation of $S_{(1)}$ has, in the massless quark
limit, a nonvanishing solution for $[\gamma_5,S_{(1)}]_+^{}$, then this 
is also a solution of the bound state equation of $\Phi_{(1)}$ for a 
pseudoscalar state with zero total momentum $P$, since $S_{(1)}$ does 
not depend on $P$. This is possible only if $P^2=0$; this means that the 
pseudoscalar state is massless and represents a Goldstone boson for 
chiral symmetry breaking.
\par 
A few comments are in order here. First, the above results are exact
statements of QCD, although for unrenormalized quantities, since no
approximations were made throughout the calculations; all expressions of 
the kernels are explicitly known in terms of Wilson loop averages and 
functional derivatives. Second, one may adopt, at the practical level
of resolution of the equations, an approximation scheme based mainly on 
the truncation of the series of kernels according to the number of sides 
of the polygonal contours of the Wilson loops. There are indications that 
the series expansions in the right-hand sides of the
integro\-diffe\-ren\-tial equations are perturbative with respect to the 
inverse of the number of sides of polygonal contours. Actually, the first
kernels $K_1$ and $N_1$ vanish for symmetry reasons and therefore the 
leading kernels are $K_2$ and $N_2$, corresponding to triangular Wilson
loops. In any event, to maintain the consistency between the various
equations, one should adopt the same type of approximation in all equations.
Third, the conclusions that are drawn are gauge invariant, since all
quantities under consideration are individually gauge invariant and this 
property is not altered by the approximation scheme of the truncation type. 
This is one of the advantages of the gauge invariant formalism.
\par

\section{Ward-Takahashi identities} \lb{s5}

Ward--Takahashi identities (WTI) express in compact form relations 
among Green's functions that emerge from symmetry properties of a 
theory. Here we sketch the WTI adapted to the gauge invariant 
formalism that we are using and corresponding to chiral symmetry.
\par
Considering the case of two different quarks, 1 and 2, we define the 
axial-vector current $j_{\mu 5}^{12}$ and the pseudoscalar density 
$v^{12}$ as
\bea \lb{5e1}
\lefteqn{\hspace{-0.6 cm}j_{\mu 5}^{12}(y)=\overline\psi_1(y)
\gamma_{\mu}\gamma_5\psi_2(y),
\ \ \ \ v^{12}(y)=i\overline\psi_1(y)\gamma_5\psi_2(y),}\nonumber \\
& &
\eea
and introduce the full axial-vector and pseudoscalar vertex functions
\bea \lb{5e2}
\lefteqn{\hspace{-0.6 cm}J_{\mu 5,\alpha\beta}^{12}(x_1,x_2;y)=
-\frac{1}{N_c}\langle 
\overline\psi_{2\beta}(x_2)U(x_2,x_1)\psi_{1\alpha}(x_1)
j_{\mu 5}^{12}(y)\rangle,}\nonumber \\
& &
\eea
\bea \lb{5e3}
\lefteqn{\hspace{-0.6 cm}P_{\alpha\beta}^{12}(x_1,x_2;y)=
-\frac{1}{N_c}\langle
\overline\psi_{2\beta}(x_2)U(x_2,x_1)\psi_{1\alpha}(x_1)v^{12}(y)
\rangle.}\nonumber \\
& &
\eea
($\alpha,\beta$ are the quark spinor indices.)
The WTI takes the form
\bea \lb{5e4}
\lefteqn{\hspace{-0.6 cm}\partial_y^{\mu}J_{\mu 5}^{12}(x_1,x_2;y)=
(m_1+m_2)P^{12}(x_1,x_2;y)}\nonumber \\
& &\hspace{-1.25 cm}-\gamma_5S_{2(1)}(x_1,x_2)\delta^4(y-x_1)
-S_{1(1)}(x_1,x_2)\gamma_5\delta^4(y-x_2).\nonumber \\
& &
\eea
\par
The analysis that follows is then standard. One passes to momentum space,
where only two momenta are independent; the total momentum $P$ enters in
the singularities of the vertex functions coming from the mesonic sector.
One first considers the chiral limit $m_1=m_2=0$, which removes the 
first term of the right-hand side of Eq. (\rf{5e4}) and transforms, in the 
limit $P\rightarrow 0$, the contributions of the 2PGIQGFs into a term 
proportional to the anticommutator of $S_{(1)}$ with $\gamma_5$. If the 
latter contribution does not vanish in these limits, then it should be 
compensated by the left-hand side. This is possible only if the 
axial-vector vertex function possesses a pseudoscalar pole at $P^2=0$.
The residue of this pole is proportional to the wave function of the state 
and to the weak decay constant $F_P$ of the Goldstone boson. One thus
finds the proportionality relation between the Goldstone boson wave 
function at $P=0$ and the scalar part of the 2PGIQGF.
\par
In the next step, one considers the case where the quark masses are
different from zero and takes the limit of $P^2$ to the physical mass
of the would-be Goldstone boson. In this limit, only the two vertex
functions of Eq. (\rf{5e4}) survive. One then considers an expansion in
the pseudoscalar vertex residue in terms of the quark masses, keeping only 
the leading contribution. Remembering that the Goldstone boson wave function
is proportional to the scalar part of the 2PGIQGF, one ends up with the
Gell-Mann--Oakes--Renner relation \cite{GellMann:1968rz}
\be \lb{5e5}
m_P^2F_P^2=-(m_1+m_2)<\overline q q>_0,
\ee
where $<\overline q q>_0$ is the quark condensate in the chiral limit
per flavor unit.    
\par

\section{Two-dimensional QCD} \lb{s6}

Two-dimensional QCD in the large-$N_c$ limit is a simplified laboratory
for real QCD \cite{'tHooft:1973jz,'tHooft:1974hx}. It possesses the
main characteristics of confinement and allows the study of the
infrared behavior of the theory in a more explicit way. On the other 
hand, Wilson loop averages can be explicitly calculated: they satisfy
the area law, due to the fact that the coupling constant of the theory
is dimensionful from the start \cite{Kazakov:1980zi,Kazakov:1980zj,
Bralic:1980ra}.
\par
Equation (\rf{2e3}) has also its counterpart in two dimensions. It 
turns out that the kernels $K_n$ for $n>2$ vanish due to the particular 
structure of the Wilson loop averages. ($K_1$ vanishes in general for 
symmetry reasons.) One remains solely with the kernel $K_2$. Equation 
(\rf{2e3}) then reduces to a nonlinear nonlocal equation in $S_{(1)}$. 
The equation can be solved explicitly and analytically by analyzing the 
singularity structure of the 2PGIQGF \cite{Sazdjian:2010ku}.
\par
The solution is infrared finite with singularities in momentum space
lying on the positive real axis of $p^2$ (timelike region). They are
represented by an infinite number of branch points, characterized by 
positive masses $M_n$ ($n=1,2,\ldots\ )$ with threshold singularities
equal to $-3/2$ in $M_n^2$. The expression of $S_{(1)}$ is, in momentum
space,
\be \lb{6e1}
S_{(1)}(p)=-i\frac{\pi}{2\sigma}\sum_{n=1}^{\infty}\,b_n
\frac{(\gamma.p+(-1)^{n+1}M_n)}{(M_n^2-p^2)^{3/2}},
\ee
where $\sigma$ is the string tension. The masses $M_n$ are greater than 
the free quark mass $m$ and ordered according to increasing values. For 
massless quarks they remain positive. The masses $M_n$ and the coefficients 
$b_n$, the latter being also positive, satisfy, for general $m$, an infinite 
set of algebraic equations that are solved numerically. Their asymptotic 
values, for large values of $n$ such that $n\gg m^2/(\pi\sigma)$, are       
\be \lb{6e2}
M_n^2\simeq \pi n\sigma,\ \ \ \ \ \   
b_n\simeq \frac{\sigma^2}{M_n+(-1)^nm}.
\ee
\par
The masses $M_n$ can be considered as dynamically generated quark masses,
since they do not exist in the QCD Lagrangian. The gauge invariant 
treatment of the problem has displayed two main features: First, the 
dynamically generated masses are infinite in number. Second, they do not 
produce simple poles in the quark Green's functions, but stronger
singularities; this fact might prevent quarks from being observed as 
asymptotic free states.    
\par
The appearance of dynamically generated masses is also accompanied by a
nonvanishing value of the quark condensate in the chiral limit. Since
the quark condensate is a local gauge invariant quantity, it can also be
calculated in various gauges. Calculations done in the light cone gauge
\cite{Zhitnitsky:1985um,Burkardt:1995eb}, 
in the axial gauge \cite{Li:1986gf} and in the present approach
\cite{Sazdjian:2010ku}, give numerically the same result. 
\par     

\section{Conclusion} \lb{s7}

Use of a gauge invariant formalism has confirmed the general relationship
that exists in QCD between the wave function of the pseudoscalar Goldstone
boson of chiral symmetry breaking and the scalar part of the two-point
gauge invariant quark Green's function, in case the latter is nonvanishing 
in the chiral limit. The relationship also persists under truncation 
schemes of the exact interaction kernels.
\par
Two-dimensional QCD in the large-$N_c$ limit, provides an explicit 
illustration of the above phenomenon. The dynamical quark mass generation
produces an infinite number of branch point singularities, stronger than
simple poles, which are infrared finite and describe the mechanism of
chiral symmetry breaking in that theory.
\par   




\bibliographystyle{elsarticle-num}
\bibliography{rdqmg-arx}







\end{document}